\documentclass[lettersize,journal,onecolumn]{IEEEtran}
\usepackage{amsmath,amsfonts}
\usepackage{array}
\usepackage[caption=false,font=normalsize,labelfont=sf,textfont=sf]{subfig}
\usepackage{textcomp}
\usepackage{url}
\usepackage{graphicx}
\usepackage{cite}
\usepackage[svgnames]{xcolor}

\newtheorem{prop}{Proposition}
\newtheorem{thm}{Theorem}
\newtheorem{lem}{Lemma}
\newtheorem{ex}{Example}

\newcommand{\ket}[1]{|#1\rangle}
\newcommand{\symp}[2]{(#1|#2)}
\newcommand{\indexset}[1]{[#1]}
\newcommand{\quantparam}[1]{[\![#1]\!]}

\DeclareMathOperator{\tr}{tr}

\begin{document}

\title{Deflating quantum error-correcting codes}
\author{Jaron Skovsted Gundersen,
    René Bødker Christensen, \\ Petar Popovski,~\IEEEmembership{Fellow, IEEE}, and Rafał Wisniewski,~\IEEEmembership{Senior Member, IEEE}.
    \thanks{J.S. Gundersen (jaron@es.aau.dk),  P. Popovski (petarp@es.aau.dk), and R. Wisniewski (raf@es.aau.dk) are with the Department of Electronic Systems at Aalborg University.}
    \thanks{R.B. Christensen (rene@math.aau.dk) is with the Department of Mathematical Sciences and the Department of Electronic Systems, both at Aalborg University.}
    }

\markboth{Article manuscript}%
{Gundersen \MakeLowercase{\textit{et al.}}: }

\maketitle

\begin{abstract}
  In this work, we introduce a technique for reducing the length of a quantum stabilizer code, and we call this \emph{deflation} of the code. Deflation can be seen as a generalization of the well-known puncturing and shortening techniques in cases where more than a single qudit is removed.
  We show that the parameters of the deflated quantum code can be controlled, and argue that a similar approach is not as beneficial when applied to classical linear codes.
  Furthermore, it is shown that deflation introduces additional freedom compared to applying just puncturing and shortening consecutively.
  We exemplify that it is possible to obtain better parameters by deflating a code rather than consecutively using puncturing and shortening.
\end{abstract}


\section{Introduction}
\IEEEPARstart{T}{wo} central challenges in coding theory are the construction of error-correcting codes with favorable parameters, and the determination of optimal bounds on achievable code parameters. Narrowing the gap between the known constructions and the theoretical limits remains a key research objective, see \cite{Grassl:codetables} for an up-to-date overview of the best known bounds and constructions.

While many codes are built from algebraic or combinatorial principles, another powerful strategy involves transforming existing codes via well-established operations such as shortening and puncturing. These techniques enable the derivation of new codes by modifying the length, dimension, and/or minimum distance. In the classical setting, shortening and puncturing are well understood and widely used to explore the trade-offs between code parameters, see for instance \cite{MacWilliamsSloane,huffman2003fundamentals}.

In the quantum setting, these operations become more nuanced. The first suggestion for puncturing stabilizer codes was discussed in \cite{CodesF4} but it was only presented for qubits and had still unexplored possibilities in terms of the qubit it was punctured with respect to. A puncturing for general qudit quantum codes was presented in \cite{Grassl21}, but as it was shown in \cite{Our_puncture}, the special case of stabilizer codes allows more flexibility in the puncturing procedure. In turn, this flexibility led to improved parameters compared to the best known codes.

Shortening has also been considered for instance in \cite{CodesF4} and \cite{Nonbinary_Stabilizer}, and the connection between a classical shortening and a quantum puncturing is discussed in \cite{Our_puncture}. 

In this work, we elaborate on the connection between shortening and puncturing which was noticed in \cite{Our_puncture}. We present a new generalized framework that encompasses both shortening and puncturing of quantum stabilizer codes, and we call this generalization \textit{deflation}. Our approach is based on the symplectic representation of the stabilizer codes. We introduce deflation, which allows fine-grained control over how and where qudits are eliminated from the code, leading to a richer set of transformation strategies than previously described. 

As an example, consider the case that we want to remove two qubits from a stabilizer code and increase the dimension by one. This can be done in $6$ different ways using a single puncturing and a single shortening (we can decide which position we puncture and which we shorten and we have three different options to choose from when puncturing). However, as we show in this paper, we increase the possibilities from $6$ to $15$ when using deflation, and the possibilities increase even more if we considered qudits. If we considered $q=3$ instead we would increase the possibilities from $8$ to $40$. The more freedom introduced by deflation opens up for the possibility that we can obtain better parameters using deflation than consecutively using puncturing and shortening, and we include an example which exactly illustrates this. 

Another natural question when deflation is introduced is if it makes sense to consider deflation in classical error correction. We discuss this in Section \ref{sec:deflationClassical}, where we argue that deflation does not appear to be as natural an operation on classical codes since the parameters seem difficult to control.

\section{Preliminaries}\label{sec:preliminaries}
Throughout this paper, let $p$ be a prime, and let $q=p^r$ for some $r\in\mathbb{Z}_+$. Denote by $\mathbb{F}_q$ the finite field of size $q$, and define the notation $\indexset{n}=\{1,2,\ldots, n\}$. For a subset $I\subseteq [n]$ we denote its complement by $\bar{I}=[n]\setminus I$.

\subsection{Stabilizer codes}\label{sec:stabilizer-codes}
To make the new presentation self-contained, we will give the necessary definitions in order to define and work with stabilizer codes. For a more in depth introduction, we refer to \cite{Ball_2023,Nonbinary_Stabilizer,gottesman1997stabilizer}.

The state of a quantum system with $q$ energy levels is called a qudit and can be represented by an element of $\mathbb{C}^q$. We fix an orthonormal basis $\{\ket{i}\mid i\in\mathbb{F}_q\}$ of $\mathbb{C}^q$ labelled by the elements of $\mathbb{F}_q$. With this notation, a possible basis of all one-qudit errors is formed by the bit flip and phase shift operators along with their products. The bit flips $X(a)$ and phase shifts $Z(b)$ are given by
\begin{align*}
  X(a)&\colon \ket{i}\mapsto\ket{i+a}\quad\text{and}\\
  Z(b)&\colon \ket{i}\mapsto\omega^{\tr(ib)}\ket{i},
\end{align*}
respectively, where $\omega\in\mathbb{C}$ is a primitive $p$'th root of unity, and $\tr$ is the (absolute) trace from $\mathbb{F}_q$ to $\mathbb{F}_p$.

Moving to multi-qudit systems, we let $\mathbf{a},\mathbf{b}\in\mathbb{F}_q^n$ and define the operators $X(\mathbf{a})=\bigotimes_{i=1}^n X(a_i)$ and similarly for $Z(\mathbf{b})$.
Then the set
\begin{equation*}
  \mathcal{E}_n = \{ X(\mathbf{a})Z(\mathbf{b}) \mid \mathbf{a},\mathbf{b}\in\mathbb{F}_q^n \}
\end{equation*}
constitutes a basis for the set of errors on an $n$-qudit system. This generates the multiplicative group
\begin{equation*}
  \mathcal{G}_n = \{ \omega^c X(\mathbf{a})Z(\mathbf{b}) \mid \mathbf{a},\mathbf{b}\in\mathbb{F}_q^n, c\in\mathbb{F}_p \},
\end{equation*}
or in the case of even characteristic (i.e. $p=2$)
\begin{equation*}
  \mathcal{G}_n = \big\{ i^c X(\mathbf{a})Z(\mathbf{b}) \mid \mathbf{a},\mathbf{b}\in\mathbb{F}_q^n, c\in\{0,1,2,3\} \big\},
\end{equation*}
where $i^2=-1$. Since the global phase of a quantum state cannot be measured, it is common to focus on the quotient $\mathcal{G}_n/\langle \omega I\rangle$ (or $\mathcal{G}_n/\langle iI\rangle$ for even characteristic), where the $I$ here indicates the identity operator, obtained by setting $c=0$, $\mathbf{a}=\mathbf{0}$, and $\mathbf{b}=\mathbf{0}$.
This quotient is isomorphic to the additive group $\mathbb{F}_q^{2n}$ via the isomorphism
\begin{equation*}
  \varphi\colon\left\{
    \begin{array}{r@{}c@{}l}
      \mathcal{G}_n/\langle\omega I\rangle & {}\rightarrow{} & \mathbb{F}_q^{2n}\\
      \omega^cX(\mathbf{a})Z(\mathbf{b})\cdot \langle\omega I\rangle & \mapsto & \symp{\mathbf{a}}{\mathbf{b}}
    \end{array}
  \right.
\end{equation*}
for $q=p^r$ with $p>2$. The case $p=2$ is similar, but with $i^c$ replacing $\omega^c$.

Since we assume that errors occur independently on qudits and that all errors are equally likely, we describe the severity of an error by using the symplectic weight $w_s$ of its corresponding element of $\mathbb{F}_q^{2n}$:
\begin{equation*}
  w_s\symp{\mathbf{a}}{\mathbf{b}} = |\{i\in\indexset{n} \mid a_i\neq 0\vee b_i\neq 0 \}|.
\end{equation*}

An $\quantparam{n,k}_q$ quantum error-correcting code is a $q^k$-dimensional subspace of $\mathbb{C}^{q^n}\simeq \mathbb{C}^q\otimes\cdots\otimes\mathbb{C}^q$, which can be interpreted as encoding $k$ logical qudits into $n$ physical qudits.

Often, such a code will be described using the stabilizer formalism. This means that the code is a subspace $\mathcal{C}_S\subseteq \mathbb{C}^{q^n}$ associated with an Abelian subgroup $S\subseteq\mathcal{G}_n$ that intersects trivially with the center of $\mathcal{G}_n$.
More precisely, the code $\mathcal{C}_S$ is given by the joint $+1$-eigenspace of the elements of $S$; that is,
\begin{equation*}
  \ket{\psi}\in\mathcal{C}_S
  \quad\Longleftrightarrow\quad
  \forall M\in S\colon M\ket{\psi} = \ket{\psi}.
\end{equation*}
If $q=p^r$, and $S$ has $r(n-k)$ independent generators, then the stabilizer code has dimension $q^n/p^{r(n-k)}=q^k$, and thus we obtain an $\quantparam{n,k}_q$ stabilizer code.

We define the map $\tilde\varphi\colon\mathcal{G}_n\rightarrow\mathbb{F}_q^{2n}$ which maps to the same as $\varphi$ but takes inputs from $\mathcal{G}_n$ instead of the equivalence classes. Furthermore, we define $S_q=\tilde\varphi(S)$ and hence a stabilizer can be described as an $\mathbb{F}_p$-linear subspace in $\mathbb{F}_q^{2n}$. We will often refer to $S_q$ as the stabilizer and $\mathcal{C}_{S_q}$ as the corresponding code.

If $S=\langle M_1,\ldots, M_{r(n-k)}\rangle$ and all the generators $M_i$ are independent, it can be shown that $\{\tilde\varphi(M_i)\}$ is $\mathbb{F}_p$-linearly independent and forms a basis of $S_q$. In $S_q$, the Abelian property of $S$ is equivalent to the symplectic product between any two vectors being $0$. The symplectic product between $\symp{\mathbf{a}}{\mathbf{b}}$ and $\symp{\mathbf{c}}{\mathbf{d}}$ in $\mathbb{F}_q^{2n}$ is given by
\begin{align}\label{eq:symplecticForm}
    \langle\symp{\mathbf{a}}{\mathbf{b}},\symp{\mathbf{c}}{\mathbf{d}}\rangle_s=\tr(\langle \mathbf{a},\mathbf{d}\rangle_E-\langle \mathbf{c},\mathbf{b}\rangle_E),
\end{align}
where $\langle\cdot,\cdot\rangle_E$ is the usual Euclidean inner product. Note that the symplectic product is an element of $\mathbb{F}_p$, and in the case where $q=p$ we can omit the trace.

Since $M\ket{\psi}=\ket{\psi}$ for any $\ket{\psi}\in \mathcal{C}_S$ and $M\in S$, an error in $S$ has no effect, and is hence `corrected' automatically in a stabilizer code. The full set of correctable errors may be determined from the Knill-Laflamme conditions~\cite{KnillLaflamme}. From this condition, the minimum distance $d$ of a stabilizer code $\mathcal{C}_S$ is
defined as 
\begin{equation}\label{eq:min_dist}
    d = \min\{w_s\symp{\mathbf{a}}{\mathbf{b}}\mid \symp{\mathbf{a}}{\mathbf{b}}\in S_q^{\perp_s}\setminus S_q\},
\end{equation}
where $S_q^{\perp_s}$ is the symplectic dual of $S_q$ with respect to the symplectic product defined in~\eqref{eq:symplecticForm}.
When the minimum distance of a code is known, we call it an $\quantparam{n,k,d}_q$ code. 
When $S_q^{\perp_s}$ contains an element of weight strictly smaller than $d$, that element is also in $S_q$, and the code is called impure. If no such element exists, the code is called pure. For additive codes, the terms pure and impure correspond exactly to the terms non-degenerate and degenerate, respectively~\cite{CodesF4}.

We can build a so-called stabilizer matrix $G_{S_q}$ by letting its $i$'th row be $\tilde\varphi(M_i)$.
In addition, the stabilizer matrix may be extended such that its rows generate $S_q^{\perp_s}$. More precisely, $S$ is Abelian so $S_q\subseteq S_q^{\perp_s}$, which implies that $2rk$ rows denoted by $G_{S_q}^{\perp_s}$ may be added to $G_{S_q}$ to obtain
\begin{equation*}
  G=\left[\begin{array}{c}
    G_{S_q}\\
    \hline
    G_{S_q}^{\perp_s}\rule{0em}{1.1em}
  \end{array}\right]\in \mathbb{F}_p^{r(n+k)\times 2n}.
\end{equation*}
Then the rows of $G$ generate $S_q^{\perp_s}$ as claimed, and we call this the extended stabilizer matrix of $S$. Note that the extended stabilizer matrix is just the stabilizer matrix for the symplectic dual. 

For each set of indices $I\subseteq\indexset{n}$, we define a projection $\pi_I\colon\mathbb{F}_q^{2n}\rightarrow \mathbb{F}_q^{2(n-|I|)}$ given by
\begin{equation}\label{eq:pi_I}
  \pi_I\big(\symp{\mathbf{a}}{\mathbf{b}}\big) = \symp{(a_j)_{j\notin I}}{(b_j)_{j\notin I}}.
\end{equation}

We also define another map needed for our deflating procedure. Let $A=\{\symp{\boldsymbol{\alpha}_1}{\boldsymbol{\beta}_1},\ldots,\symp{\boldsymbol{\alpha}_\ell}{\boldsymbol{\beta}_\ell}\}$ consist of elements in $\mathbb{F}_q^{2t}$ for some $t<n$, and let $I\subseteq [n]$ with $|I|=t$. Then, we define
\begin{equation}\label{eq:sigma}
    \sigma_{A,I}\colon\left\{
    \begin{array}{r@{}c@{}l}
      \mathbb{F}_q^{2n} & \rightarrow & \mathbb{F}_p^{\ell}\\
      \symp{\mathbf{a}}{\mathbf{b}} & \mapsto & (\langle \pi_{\bar{I}}\symp{\mathbf{a}}{\mathbf{b}},\symp{\boldsymbol{\alpha}_j}{\boldsymbol{\beta}_j}\rangle_s)_{j=1,\ldots,\ell}
    \end{array}
  \right. .
\end{equation}
Note that $\pi_{\bar{I}}(\mathrm{ker} \sigma_{A,I})$ is exactly $\mathrm{span}_{\mathbb{F}_p}(A)^{\perp_s}$.
\section{Deflation: A generalization of shortening and puncturing }\label{sec:deflation}
Puncturing and shortening error correcting codes are described both in the classical and quantum setting. However, as shown in \cite{Our_puncture}, it is possible to introduce more freedom into the puncturing of a stabilizer codes, which has led to improved parameters for the best known quantum codes. In this section, we will make a general framework for constructing new codes with fewer qudits than the original code. We call this framework deflation, and it includes the puncturing and shortening techniques as special cases. In this way, we introduce even more freedom into the puncturing and shortening techniques. 

The codes constructed using our proposed technique are defined using the maps in \eqref{eq:pi_I} and \eqref{eq:sigma}. We start with an $\quantparam{n,k}$ code $\mathcal{C}_{S_q}$ and introduce a shorter $\quantparam{t,k'}$ code $\mathcal{C}_{S_q'}$. Now let $B_{S_q'}$ be a basis of $S_q'^{\perp_s}$. We define the deflated code of $\mathcal{C}_{S_q}$ with respect to $\mathcal{C}_{S_q'}$ on the indices in $I$ by the stabilizer
\begin{equation}\label{eq:def_puncturing}
    S_q^{S_q',I} = \pi_I(\mathrm{ker}(\sigma_{B_{S_q'},I})).
\end{equation}
Notice that an element in the kernel of $\sigma_{B_{S_q'},I}$ are vectors that are mapped to $S_q'$ under $\pi_{\bar{I}}$. Applying $\pi_I$ removes exactly this part of the vector, which we will call the \textit{prefix}. So to explain this new stabilizer, we first restrict ourselves to the elements of $S_q$ having prefix in $S_q'$, before deleting these entries. 

We first prove that $S_q^{S_q',I}$ is indeed a stabilizer. In order to ease the notation, we will often write $\symp{\mathbf{a},\mathbf{u}}{\mathbf{b},\mathbf{v}}$ for an element in $S_q$, where we assume that $\pi_I\symp{\mathbf{a},\mathbf{u}}{\mathbf{b},\mathbf{v}}=\symp{\mathbf{u}}{\mathbf{v}}$ and $\pi_{\bar{I}}\symp{\mathbf{a},\mathbf{u}}{\mathbf{b},\mathbf{v}}=\symp{\mathbf{a}}{\mathbf{b}}$. This will of course only be the case if $I=\indexset{t}$, but we remark that reordering the qudits results in an equivalent code, so we could actually w.l.o.g. assume that $I=\indexset{t}$. This is also the reason why, we often drop the $I$ superscript in \eqref{eq:def_puncturing} and simply write $S_q^{S_q'}$
\begin{prop}
    $S_q^{S_q'}$ is a stabilizer.
\end{prop}
\begin{IEEEproof}
  We need to show that any pair of elements in $S_q^{S_q'}$ have a symplectic product equal to $0$. More formally, for any $\symp{\mathbf{u}}{\mathbf{v}},\symp{\mathbf{u}'}{\mathbf{v}'}\in S_q^{S_q'}$ we can find prefixes $\symp{\mathbf{a}}{\mathbf{b}}$ and $\symp{\mathbf{a}'}{\mathbf{b}'}$ of $S_q'$ such that $\symp{\mathbf{a},\mathbf{u}}{\mathbf{b},\mathbf{v}}\in S_q$. The fact that $S_q'$ is a stabilizer implies that
  \begin{align*}
    0 &= \langle \symp{\mathbf{a},\mathbf{u}}{\mathbf{b},\mathbf{v}}, \symp{\mathbf{a}',\mathbf{u}'}{\mathbf{b}',\mathbf{v}'}\rangle_s\\
      &= \langle\symp{\mathbf{a}}{\mathbf{b}}, \symp{\mathbf{a'}}{\mathbf{b}'}\rangle_s + \langle\symp{\mathbf{u}}{\mathbf{v}}, \symp{\mathbf{u}'}{\mathbf{v}'}\rangle_s\\
    &= \langle\symp{\mathbf{u}}{\mathbf{v}}, \symp{\mathbf{u}'}{\mathbf{v}'} \rangle_s,
  \end{align*}
  as desired.
\end{IEEEproof}
It turns out that the dual of this deflated stabilizer can also be obtained by deflating the dual of the original stabilizer.
\begin{prop}\label{prop:dual}
  Let $\mathcal{C}_{S_q}$ be an $\quantparam{n,k,d}_q$ stabilizer code, and $\mathcal{C}_{S_q'}$ a $\quantparam{t,k'}_q$ stabilizer code with $t<d$. Then
  \begin{equation*}
    (S_q^{S_q'})^{\perp_s} = (S_q^{\perp_s})^{S_q'^{\perp_s}}.
  \end{equation*}
\end{prop}
\begin{IEEEproof}
  First, let $\symp{\mathbf{u}}{\mathbf{v}}\in (S_q^{\perp_s})^{S_q'^{\perp_s}}$. This means that there exists a prefix $\symp{\mathbf{a}}{\mathbf{b}}\in S_q'^{\perp_s}$ such that $\symp{\mathbf{a},\mathbf{u}}{\mathbf{b},\mathbf{v}}\in S_q^{\perp_s}$.
  Now, consider an arbitrary element $\symp{\mathbf{a}',\mathbf{u}'}{\mathbf{b}',\mathbf{v}'}\in S_q$. By definition, $\langle\symp{\mathbf{a},\mathbf{u}}{\mathbf{b},\mathbf{v}}, \symp{\mathbf{a}',\mathbf{u}'}{\mathbf{b}',\mathbf{v}'}\rangle_s =0$. If $\symp{\mathbf{a}'}{\mathbf{b}'}\in S_q'$, we have $\langle\symp{\mathbf{a}}{\mathbf{b}}, \symp{\mathbf{a}'}{\mathbf{b}'}\rangle_s=0$, which in turn implies $\langle \symp{\mathbf{u}}{\mathbf{v}}, \symp{\mathbf{u}'}{\mathbf{v}'}\rangle_s = 0$. These observations imply $\symp{\mathbf{u}}{\mathbf{v}}\in (S_q^{S_q'})^{\perp_s}$.

  For the other inclusion, let $\symp{\mathbf{u}}{\mathbf{v}}\in (S_q^{S_q'})^{\perp_s}$. We need to show the existence of a prefix $\symp{\mathbf{a}}{\mathbf{b}}\in S_q'^{\perp_s}$ such that $\symp{\mathbf{a},\mathbf{u}}{\mathbf{b},\mathbf{v}}\in S_q^{\perp_s}$.
  Now, let $\{\symp{\mathbf{a}_i,\mathbf{u}_i}{\mathbf{b}_i,\mathbf{v}_i}\}_{j=1}^{r(n-k)}$ be a basis of $S_q$, and let $\{\symp{\boldsymbol\alpha_j}{\boldsymbol\beta_j}\}_{j=1}^{r(t+k')}$ be a basis for $S_q'^{\perp_s}$.
  If the prefix $\symp{\mathbf{a}}{\mathbf{b}}\in S_q'^{\perp_s}$ exists, it must have the form
  \begin{equation*}
    \symp{\mathbf{a}}{\mathbf{b}} = \sum_{j=1}^{r(t+k')} c_j\symp{\boldsymbol\alpha_j}{\boldsymbol\beta_j},\quad c_j\in\mathbb{F}_p.
  \end{equation*}
  Thus, we obtain the result if we can find such $c_j\in\mathbb{F}_p$ that satisfy $\langle\symp{\mathbf{a},\mathbf{u}}{\mathbf{b},\mathbf{v}},\symp{\mathbf{a}_i,\mathbf{u}_i}{\mathbf{b}_i,\mathbf{v}_i},\rangle_s = 0$ for every $i\in\indexset{r(n-k)}$.
  To this end, let $\delta_{ij}=\langle\symp{\boldsymbol\alpha_j}{\boldsymbol\beta_j}, \symp{\mathbf{a}_i}{\mathbf{b}_i}\rangle_s$, and collect these in a matrix $\Delta\in\mathbb{F}_p^{r(n-k)\times r(t+k')}$.
  Further, let $\gamma_i=\langle\symp{\mathbf{u}}{\mathbf{v}}, \symp{\mathbf{u}_i}{\mathbf{v}_i}\rangle_s$ be the entries of a vector $\boldsymbol\gamma\in\mathbb{F}_p^{r(n-k)}$.
  With these notations, we are searching for a solution of the linear system
  \begin{equation*}
    \Delta\mathbf{c} = -\boldsymbol\gamma.
  \end{equation*}
  This system has a solution if and only if the reduced row-echelon form of $[\Delta\mid -\boldsymbol\gamma]$ has no pivot in the last column. For such a pivot to appear, there must exist $\mathbf{s}\in\mathbb{F}_p^{r(n-k)}$ such that $\mathbf{s}^\top\Delta=\mathbf{0}$, but $\mathbf{s}^\top(-\boldsymbol\gamma)\neq 0$. If $\mathbf{s}^\top\Delta=\mathbf{0}$, we have $\langle\symp{\boldsymbol\alpha_j}{\boldsymbol\beta_j}, \sum_i s_i\symp{\mathbf{a}_i}{\mathbf{b}_i}\rangle_s=0$ for every $j$, which implies $\sum s_i\symp{\mathbf{a}_i}{\mathbf{b}_i}\in S_q'$. But then $\sum_i s_i\symp{\mathbf{a}_i,\mathbf{u}_i}{\mathbf{b}_i,\mathbf{v}_i}\in\ker\sigma_{S_q',I}$, from which we obtain $\sum_i s_i\symp{\mathbf{u}_i}{\mathbf{v}_i}\in S_q^{S_q'}$. The assumption on $\symp{\mathbf{u}}{\mathbf{v}}$ and the construction of $\boldsymbol\gamma$ now imply $\mathbf{s}^\top\boldsymbol\gamma = 0$, and no such pivot can exist.
  In conclusion, we can find the $c_j$, and hence the prefix $\symp{\mathbf{a}}{\mathbf{b}}$, as desired.
\end{IEEEproof}

Having now established that $S_q^{S_q'}$ is indeed a stabilizer, we relate this technique to both shortening and puncturing. A shortening at a single entry is carried out by removing all the vectors from $S_q$ not having $(0|0)$ on this entry and afterwards removing this entry. This can easily be generalized to multiple entries where all vectors with $\pi_{\bar{I}}\symp{\mathbf{a}}{\mathbf{b}}\neq \symp{\mathbf{0}}{\mathbf{0}}$ are removed from $S_q$, before removing these $0$-entries from the surviving vectors. Note, however, that we can capture this in our technique by letting $S_q'=\{\symp{\mathbf{0}}{\mathbf{0}}\}$.

Similarly, we can recover the puncturing technique introduced in~\cite{Our_puncture} as a special case. Namely, a puncturing at a single entry is carried out by fixing $r$ $\mathbb{F}_p$-linear independent elements $\symp{\alpha^{(1)}}{\beta^{(1)}},\ldots,\symp{\alpha^{(r)}}{\beta^{(r)}}$ and then deleting all vectors from $S_q$ which on this entry is not in $\mathrm{span}(\{\symp{\alpha^{(1)}}{\beta^{(1)}},\ldots,\symp{\alpha^{(r)}}{\beta^{(r)}}\})$.
If this puncturing is repeated $t$ times, it is equivalent to restricting the prefixes to the span of 
\begin{align}\label{eq:puncture_kernel}
  \begin{aligned}
    \{(\alpha_1^{(1)},0,0,\ldots,0&|\beta_1^{(1)},0,0\ldots,0),\\&\vdots\\(\alpha_1^{(r)},0,0,\ldots,0&|\beta_1^{(r)},0,0\ldots,0),\\
      (0,\alpha_2^{(1)},0,\ldots,0&|0,\beta_2^{(1)},0\ldots,0),\\&
      \vdots\\
      (0,\alpha_2^{(r)},0,\ldots,0&|0,\beta_2^{(r)},0\ldots,0),\\&\vdots\\
      (0,\ldots,0,0,\alpha_t^{(1)}&|0,\ldots,0,0,\beta_t^{(1)}).\\&\vdots\\
      (0,\ldots,0,0,\alpha_t^{(r)}&|0,\ldots,0,0,\beta_t^{(r)})\},
    \end{aligned}
\end{align}
which corresponds exactly to choosing $S_q'$ to be the span of \eqref{eq:puncture_kernel}. 

For now, it is not clear that we gain something by introducing deflation. However, after determining what the parameters of the deflated code is, we will in Section \ref{sec:DeflateVsPandS} show that deflation introduces more freedom than just consecutively performing puncturings and shortenings. 

\subsection{Parameters for pure codes}\label{sec:pure-codes}
We now determine the dimension and minimum distance of $S_q^{S_q'}$ when $\mathcal{C}_{S_q}$ is a pure code, meaning that the symplectic weight of elements in $S_q$ is at least $d$. First, we need the following lemma.
\begin{lem}\label{lem:firstIndicesDimPure}
   Let $\mathcal{C}_{S_q}$ be a pure $\quantparam{n,k,d}_q$ stabilizer code and $I\subseteq \indexset{n}$ with $|I|=t$. If $t<d$, then $\dim(\pi_{\bar{I}}(S_q))=2rt$.
\end{lem}
\begin{IEEEproof}
  Assume that $\dim(\pi_{\bar{I}}(S_q))<2rt$. This implies the existence of a non-zero $\symp{\mathbf{a}}{\mathbf{b}}\in (\pi_{\bar{I}}(S_q))^{\perp_s}$ and therefore $\symp{\mathbf{a},\mathbf{0}}{\mathbf{b},\mathbf{0}}\in S_q^{\perp_s}$. But this contradicts the fact that $S_q$ is pure of minimal distance $d$.
\end{IEEEproof}

\begin{thm}\label{thm:parametersPure}
  Let $\mathcal{C}_{S_q}$ be a pure $\quantparam{n,k,d}_q$ stabilizer code, $\mathcal{C}_{S_q'}$ be a $\quantparam{t,k'}_q$ stabilizer code, and $I\subseteq \indexset{n}$ with $|I|=t$. If $t<d$, then $\mathcal{C}_{S_q^{S_q'}}$ is an $\quantparam{n-t,k+k',\tilde{d}}$ stabilizer code with $\tilde{d}\geq d-t$.
\end{thm}
\begin{IEEEproof}
  Clearly, $\mathcal{C}_{S_q^{S_q'}}$ has length $n-t$ since $t$ indices are removed.
  To determine the dimension, note that $\dim S_q' = r(t-k')$. Lemma~\ref{lem:firstIndicesDimPure} now implies that $S_q$ has a basis on the form $\mathcal{B}_1\cup\mathcal{B}_2\cup\mathcal{B}_3$ where
  \begin{align*}
    \mathcal{B}_1 &= \{\symp{\mathbf{a}_j,\mathbf{u}_j}{\mathbf{b}_j,\mathbf{v}_j}\}_{j=1}^{r(t-k')}\\
    \mathcal{B}_2 &= \{\symp{\mathbf{a}_j,\mathbf{u}_j}{\mathbf{b}_j,\mathbf{v}_j}\}_{j=r(t-k')+1}^{2rt}\\
    \mathcal{B}_3 &= \{\symp{\mathbf{0},\mathbf{u}_j}{\mathbf{0},\mathbf{v}_j}\}_{j=2rt+1}^{r(n-k)}.
  \end{align*}
  such that $\pi_{\bar I}(\mathcal{B}_1)$ is a basis of $S_q'$, and $\pi_{\bar I}(\mathcal{B}_1\cup\mathcal{B}_2)$ is a basis of $\pi_{\bar{I}}(S_q)$. By construction of $S_q^{S_q'}$, it is spanned by $\pi_I(\mathcal{B}_1\cup\mathcal{B}_3)$. In fact, this is a basis since it would otherwise contradict $\mathcal{B}_1\cup\mathcal{B}_2\cup\mathcal{B}_3$ being a basis or $\mathcal{C}_{S_q}$ being pure.
  Thus, the dimension of $S_q$ is
  \begin{equation*}
    r(t-k') + \big(r(n-k)-2rt\big) = r\big(n-t-(k+k')\big),
  \end{equation*}
  yielding the claimed dimension of $\mathcal{C}_{S_q^{S_q'}}$. 
  
  For the minimum distance we need to find the minimum symplectic weight of elements in $(S_q^{S_q'})^{\perp_s}\setminus S_q^{S_q'}$. According to Proposition \ref{prop:dual}, $(S_q^{S_q'})^{\perp_s}=(S_q^{\perp_s})^{S_q'{\perp_s}}$, so all the elements we need to consider are some of the elements from $S_q^{\perp_s}$ with $t$ entries removed. Since $\mathcal{C}_{S_q}$ is pure, all elements in $S_q^{\perp_s}$ have weight at least $d$, implying that the deflated code has minimum distance at least $\tilde d \geq d-t$.
\end{IEEEproof}
We once again highlight its relation to puncturing and shortening.
When successively shortening a pure stabilizer code in $t$ positions, the dimension will increase by $t$, and only those symplectic vectors with prefix $\symp{\mathbf{0}}{\mathbf{0}}$ remain. As described earlier, this corresponds to $S_q'$ containing only the zero vector, meaning that $C_{S_q'}$ is a $\quantparam{t,t}_q$ code. Thus, the proposition gives a dimension increase of $t$ as expected.

On the other hand, successive puncturing will retain the dimension of the code, but each position $i$ will be fixed within the ($r$-dimensional) space $\mathrm{Span}_{\mathbb{F}_p}\{\symp{\alpha^{(j)}}{\beta^{(j)}}\}_{j=1}^r$. This corresponds to $S_q'$ being $rt$-dimensional, and hence defining a $\quantparam{t,0}_q$ code. Once again, this matches the dimension difference given in the proposition.

\subsection{Impure codes}\label{sec:impure-codes}
While the purity of codes is a necessary condition for shortening, this is not the case for puncturing. In this section, we consider impure codes and strengthen our assumptions to extend our deflation technique to include these codes as well.
\begin{lem}\label{lem:basis_dual_prefix}
    If $|I|=t<d$ and $\symp{\mathbf{a}}{\mathbf{b}}\in (\pi_{\bar I}(S_q))^{\perp_s}$, then $\symp{\mathbf{a},\mathbf{0}}{\mathbf{b},\mathbf{0}}\in S_q$
\end{lem}
\begin{IEEEproof}
    Clearly, if  $\symp{\mathbf{a}}{\mathbf{b}}\in (\pi_{\bar I}(S_q))^{\perp_s}$ the element $\symp{\mathbf{a},\mathbf{0}}{\mathbf{b},\mathbf{0}}$ must be in $S_q^{\perp_s}$. However, this element has weight $t<d$ and can therefore not be in $S_q^{\perp_s}\setminus S_q$. 
\end{IEEEproof}
\begin{lem}
  Let $\mathcal{C}_{S_q}$ be an $\quantparam{n,k,d}_q$ stabilizer code, and $\mathcal{C}_{S_q'}$ a $\quantparam{t,k'}_q$ stabilizer code. If $t<d$, then $(\pi_{\bar I}(S_q))^{\perp_s}\subseteq\pi_{\bar I}(S_q)$ and $rt\leq \dim\pi_{\bar I}(S_q)$.
\end{lem}
\begin{IEEEproof}
  Write $m=\dim\pi_{\bar I}(S_q)$. We then have $2rt-m$ linearly independent vectors $\symp{\mathbf{a}_1}{\mathbf{b}_1},\ldots,\symp{\mathbf{a}_{2rt-m}}{\mathbf{b}_{2rt-m}}$ that span $(\pi_{\bar I}(S_q))^{\perp_s}$.
  Due to Lemma \ref{lem:basis_dual_prefix}, the elements $\symp{\mathbf{a}_1,\mathbf{0}}{\mathbf{b}_1,\mathbf{0}},\ldots,\symp{\mathbf{a}_{2rt-m},\mathbf{0}}{\mathbf{b}_{2rt-m}, \mathbf{0}}$ must be in $S_q$ and they are linearly independent. This shows that $(\pi_{\bar I}(S_q))^{\perp_s}\subseteq\pi_{\bar I}(S_q)$. 
  
  In terms of the dimensions, this inclusion implies $2rt-\dim\pi_{\bar I}(S_q)\leq\dim\pi_{\bar I}(S_q)$, yielding the last claim. 
\end{IEEEproof}
Our assumption from now on will be that we choose an $S_q'$ such that $(S_q')^{\perp_s}\subseteq\pi_{\bar I}(S_q)$. This implies the inclusions
\begin{equation}\label{eq:inclusions}
  (\pi_{\bar I}(S_q))^{\perp_s}\subseteq S_q'
  \subseteq (S_q')^{\perp_s}\subseteq \pi_{\bar I}(S_q).
\end{equation}
We remark that for the pure codes $\dim \pi_{\bar I}(S_q)=2rt$ due to Lemma \ref{lem:firstIndicesDimPure}, meaning that it is the full space and therefore~\eqref{eq:inclusions} is trivially fulfilled. 

\begin{thm}\label{thm:parameters_impure}
  Let $\mathcal{C}_{S_q}$ be an $\quantparam{n,k,d}_q$ stabilizer code and $I\subseteq \indexset{n}$ with $|I|=t$. If $\mathcal{C}_{S_q'}$ is a $\quantparam{t,k'}_q$ stabilizer code satisfying $(S_q')^{\perp_s}\subseteq \pi_{\bar{I}}(S_q)$, then $\mathcal{C}_{S_q^{S_q'}}$ has parameters $\quantparam{n-t, k+k',\tilde d}_q$.
  Furthermore, if every element of $S_q$ with prefix in $S_q'^{\perp_s}\setminus S_q'$ has weight at least $d$, we have that $\tilde d \geq d-t$.
\end{thm}
\begin{IEEEproof}
  Again, clearly the length of $\mathcal{C}_{S_q^{S_q'}}$ is $n-t$ since $t$ entries are removed.

  To determine the dimension we use an approach similar to the proof of Theorem~\ref{thm:parametersPure}. As previously, write $m=\dim \pi_{\bar{I}}(S_q)$. By the choice of $S_q'$, the inclusions in~\eqref{eq:inclusions} hold. From this and Lemma \ref{lem:basis_dual_prefix}, $S_q$ has a basis on the form $\bigcup_{i=1}^4 \mathcal{B}_i$ with
  \begin{align*}
    \mathcal{B}_1 &= \{\symp{\mathbf{a}_j,\mathbf{0}}{\mathbf{b}_j,\mathbf{0}}\}_{j=1}^{2rt-m}\\
    \mathcal{B}_2 &= \{\symp{\mathbf{a}_j,\mathbf{u}_j}{\mathbf{b}_j,\mathbf{v}_j}\}_{j=2rt-m+1}^{r(t-k')}\\
    \mathcal{B}_3 &= \{\symp{\mathbf{a}_j,\mathbf{u}_j}{\mathbf{b}_j,\mathbf{v}_j}\}_{j=r(t-k')+1}^{m}\\
    \mathcal{B}_4 &= \{\symp{\mathbf{0},\mathbf{u}_j}{\mathbf{0},\mathbf{v}_j}\}_{j=m+1}^{r(n-k)}\\
  \end{align*}
  such that $\pi_{\bar I}(\mathcal{B}_1)$ is a basis of $(\pi_{\bar{I}}(S_q))^{\perp_s}$, $\pi_{\bar I}(\mathcal{B}_1\cup\mathcal{B}_2)$ a basis of $S_q'$, and $\pi_{\bar I}(\bigcup_{i=1}^3\mathcal{B}_i)$ a basis of $\pi_{\bar{I}}(S_q)$.

  Now, $\ker\sigma_{B_{S_q'},I}$ has basis $\mathcal{B}_1\cup\mathcal{B}_2\cup\mathcal{B}_4$.
  We claim that applying $\pi_I$ removes exactly the basis elements in $\mathcal{B}_1$. To see this, note that the dimension loss from applying $\pi_I$ comes from the elements in $\ker\pi_I\cap\ker\sigma_{B_{S_q'},I}$. Such an element has the form $\symp{\mathbf{a},\mathbf{0}}{\mathbf{b},\mathbf{0}}\in S_q$ with $\symp{\mathbf{a}}{\mathbf{b}}\in S_q'$.
  However, $\symp{\mathbf{a},\mathbf{0}}{\mathbf{b},\mathbf{0}}\in S_q$ if and only if $\langle\symp{\mathbf{a}}{\mathbf{b}}, \symp{\mathbf{a}'}{\mathbf{b}'}\rangle_s=0$ for all elements $\symp{\mathbf{a}'}{\mathbf{b}'}\in\pi_{\bar{I}}(S_q)$. By definition, this is equivalent to $\symp{\mathbf{a}}{\mathbf{b}}\in(\pi_{\bar{I}}(S_q))^{\perp_s}$.

  Thus, the deflated code is spanned by $\pi_{I}(\mathcal{B}_2\cup \mathcal{B}_4)$. This is a basis if the elements are linearly independent. Assume that they are not, meaning that there exist $c_j$'s not all zeros such that 
  \begin{equation*}
    \sum_{j=2rt-m+1}^{r(t-k')}\hspace{-4mm} c_j\pi_{\bar{I}}\symp{\mathbf{a}_j,\mathbf{u}_j}{\mathbf{b}_j,\mathbf{v}_j}
    +\!\sum_{j=m+1}^{r(n-k)}\!\! c_j\pi_{\bar{I}}\symp{\mathbf{a}_j,\mathbf{u}_j}{\mathbf{b}_j,\mathbf{v}_j}=\mathbf{0}.
  \end{equation*}
  Since the original vectors are linearly independent the same linear combination on the original vectors is nonzero meaning that
    \begin{equation*}
      \sum_{j=2rt-m+1}^{r(t-k')}c_j\symp{\mathbf{a}_j}{\mathbf{b}_j}+\sum_{j=m+1}^{r(n-k)}c_j\symp{\mathbf{a}_j}{\mathbf{b}_j}\neq\mathbf{0}.
  \end{equation*}
  But this implies that 
  \begin{equation*}
    \sum_{j=2rt-m+1}^{r(t-k')}\hspace{-5mm} c_j\symp{\mathbf{a}_j,\mathbf{u}_j}{\mathbf{b}_j,\mathbf{v}_j}
    +\!\sum_{j=m+1}^{r(n-k)}\hspace{-2mm} c_j\symp{\mathbf{a}_j,\mathbf{u}_j}{\mathbf{b}_j,\mathbf{v}_j}
    \!\in\! \mathrm{Span}(\mathcal{B}_1),
  \end{equation*}
  contradicting the fact that $\cup_{i=1}^4 \mathcal{B}_i$ is a basis.
  
  Therefore, the dimension of $S_q^{S_q'}$ is
  \begin{align*}
      r(t-k')-(2rt-m)+r(n-k)-m = r(n-t-k-k')
  \end{align*}
  implying that the stabilizer code has dimension $k+k'$.

  For the minimum distance, the assumption regarding prefixes ensure that no elements with weight less than $d$ from $S_q$ will be `moved' to the dual after deflating and hence we conclude that $\tilde d\geq d-t$.
\end{IEEEproof}

\section{Deflation versus puncturing and shortening}\label{sec:DeflateVsPandS}
In this section, we highlight the flexibility obtained by introducing deflation as a generalization of both puncturing and shortening. To illustrate this, we consider a case where $q=2$ and $t=2$. We consider an $[\![n,k,d]\!]_2$ code given by $S_2$. We want to remove two entries and increase the dimension by $1$, implying that we need an $S_2'$ defining a $\quantparam{2,1}_2$ code.
In that case, $S_2^{S_2'}$ will be an $[\![n-2,k+1,d-2]\!]_2$ code. Note that $S_2'$ is the span of a single nonzero vector in $\mathbb{F}_2^{2\cdot 2}$ and one way to obtain this code is by performing a single puncturing and a single shortening.
Since puncturing can be done wrt. $3$ different elements and at $2$ different positions, we have a total of $6$ different prefixes spanning $S_2'$ which are given by
\begin{align}\label{eq:prefix_punc_short}
    \symp{00}{01}, \symp{01}{00},\symp{01}{01},
    \symp{00}{10},\symp{10}{00},\symp{10}{10}.
\end{align}
However, the number of possible prefixes when performing deflation is $2^4-1=15$ since any single element from $\mathbb{F}_2^{2\cdot 2}\setminus\{\mathbf{0}\}$ defines a $[\![2,1]\!]_2$ code. Thus, in this fairly small example we have increased the number of potential $S_2'$ from $6$ to $15$. In the following example, we show how this potential can be used to obtain codes with improved parameters. 
\begin{ex}\label{ex:ImprovedDistance}
    Define a $\quantparam{8,1,2}_2$ stabilizer code with extended stabilizer matrix
    {\tiny
\[
\left[
\begin{array}{cccccccc|cccccccc}
    1 & 0 & 0 & 0 & 0 & 0 & 0 & 0 & 0 & 0 & 1 & 0 & 0 & 0 & 0 & 0 \\
    0 & 1 & 0 & 0 & 0 & 0 & 0 & 0 & 0 & 0 & 0 & 1 & 0 & 0 & 0 & 0 \\
    0 & 0 & 1 & 0 & 0 & 0 & 1 & 0 & 1 & 0 & 0 & 0 & 0 & 1 & 0 & 0 \\
    0 & 0 & 0 & 1 & 0 & 0 & 0 & 1 & 0 & 1 & 1 & 0 & 1 & 1 & 1 & 1 \\
    0 & 0 & 0 & 0 & 1 & 0 & 0 & 0 & 0 & 0 & 1 & 1 & 0 & 1 & 1 & 0 \\
    0 & 0 & 0 & 0 & 0 & 1 & 1 & 1 & 0 & 0 & 0 & 0 & 0 & 0 & 1 & 1 \\
    0 & 0 & 0 & 0 & 0 & 0 & 0 & 0 & 0 & 0 & 0 & 1 & 0 & 1 & 0 & 1 \\
    \hline
    0 & 0 & 0 & 0 & 0 & 0 & 1 & 0 & 0 & 0 & 0 & 1 & 1 & 1 & 0 & 0 \\
    0 & 0 & 0 & 0 & 0 & 0 & 0 & 0 & 0 & 0 & 1 & 0 & 0 & 1 & 1 & 0 
\end{array}
\right]
\]}
Now notice that deflating the first two entries with respect to one of the prefixes in \eqref{eq:prefix_punc_short} (letting $S_2'$ be the span of one of the prefixes) causes either $\symp{10}{00}$ or $\symp{01}{00}$ to be in $S_2'^{\perp_s}\setminus S_2'$.
Any element having a prefix in $S_2'^{\perp_s}\setminus S_2'$ will be moved to the dual of the deflated code, and hence the weight of these elements influence the minimum distance of the deflated code.
In either case, this holds for one of the two topmost rows in the original extended stabilizer matrix, implying that we end up with a $\quantparam{6,2,1}_2$ code. Thus, if we puncture and shorten on the first and second entry, the minimum distance will always be $1$.
However, if we deflate with respect to $S_2'=\mathrm{span}\{\symp{11}{11}\}$, the dual is $S_2^{\perp_s}=\mathrm{span}\{\symp{11}{11},\symp{10}{10},\symp{11}{00}\}$ and this contains neither $\symp{10}{00}$ nor $\symp{01}{00}$, which were the prefixes causing the problems before. Performing this deflation will actually give us a $\quantparam{6,2,2}_2$ code.
\end{ex}
To be more general, we can count the number of possibilities we have with deflation by computing the number of $\quantparam{t,k'}$ stabilizer codes. In \cite{CountingStabilizerCodes}, they prove how many stabilizer codes there exists when considering the qudits as elements over $\mathbb{Z}_q$. Since we have described qudits as elements over $\mathbb{F}_q$, we modify their proof below. The main difference is that we need to choose more generating elements for our stabilizer in order to obtain the right dimension since we consider the stabilizer space as an $\mathbb{F}_p$-dimensional subspace.
\begin{prop}\label{prop:stabCount}
    The number of $\quantparam{n,k}_q$ stabilizer codes, where $q=p^r$ is given by
    \begin{align*}
        \prod_{i=0}^{r(n-k)-1}\frac{p^{2rn-i}-p^i}{p^{r(n-k)}-p^i}
    \end{align*}
\end{prop}
\begin{IEEEproof}
    We start by counting the number of ways to choose $r(n-k)$ $\mathbb{F}_p$-linearly independent and symplectically orthogonal vectors in $\mathbb{F}_q^{2n}$. For the first vector, we are free to choose anything except the zero vector, giving us $p^{2rn}-1$ possible vectors. For the next vector, we have $p^{2rn-1}$ elements in the symplectic dual of the first vector. Furthermore, every vector is symplectically orthogonal to itself, so from the $p^{2rn-1}$ vectors, we need to subtract the $p$ vectors lying in the span of the vector chosen first. Continuing in this way, we obtain 
    \begin{align*}
         \prod_{i=0}^{r(n-k)-1}p^{2rn-i}-p^i
    \end{align*}
    ways to choose $r(n-k)$ $\mathbb{F}_p$-linearly independent and symplectically orthogonal vectors in $\mathbb{F}_q^{2n}$.

    However, some of the above choices might result in the same subspace. Hence, we need to divide by the number of possible bases for an $r(n-k)$-dimensional subsspace. We argue in a similar fashion as above. For the first vector we are free to choose any element except the zero vector in the space, giving us $p^{r(n-k)}-1$ possibilities. For the next we can choose any element not in the span of the first, giving us $p^{r(n-k)}-p$ possibilities. Continuing in this fashion gives us
    \begin{align*}
        \prod_{i=0}^{r(n-k)-1}p^{r(n-k)}-p^i
    \end{align*}
    possible bases for the subspace. In total, we conclude that the number of $\quantparam{n,k}_q$ stabilizer codes is 
    \begin{align*}
        \prod_{i=0}^{r(n-k)-1}\frac{p^{2rn-i}-p^i}{p^{r(n-k)}-p^i}.
    \end{align*}
\end{IEEEproof}
We are interested in the number of $\quantparam{t,k'}_q$ stabilizer codes since this gives us exactly the possible ways to carry out deflating on $t$ positions with a specific dimension increase $k'$. 

We can compare this to the number of possibilities we have when performing consecutive puncturings and shortenings. If we want the dimension increase to be the same as deflating with respect to a $\quantparam{t,k'}$ stabilizer code $S'$ we need to carry out $k'$ shortenings and $t-k'$ puncturings. The number of ways to carry out $t-k'$ puncturings can also be deduced from Proposition \ref{prop:stabCount} since $t-k'$ puncturings correspond to choosing a $\quantparam{t-k',0}_q$ code.
A shortening can only be performed in a single way, but we must decide which positions to shorten and which to puncture. In total, we have 
\begin{align*}
    \binom{t}{k'}\prod_{i=0}^{r(t-k')-1}\frac{p^{2r(t-k')-i}-p^i}{p^{r(t-k')}-p^i}
\end{align*}
possible ways to carry out $k'$ shortenings and $t-k'$ puncturings. To compare the number of possibilities for the different methods we have included the values for specific parameters in Tables~\ref{tab:Poss_t2k1} and \ref{tab:Poss_t3k1}.

\begin{table}[]
\centering
\begin{tabular}{l|ccc|ccc|}
\cline{2-7}
                                       &                            & $p=2$                      &       &                            & $p=3$                      &       \\ \cline{2-7} 
                                       &\multicolumn{1}{c|}{$r=1$} & \multicolumn{1}{c|}{$r=2$} & $r=3$ & \multicolumn{1}{c|}{$r=1$} & \multicolumn{1}{c|}{$r=2$} & $r=3$ \\ \hline
\multicolumn{1}{|l|}{Punc. and short.}        & \multicolumn{1}{c|}{$6$}      & \multicolumn{1}{c|}{$30$}      & $270$      & \multicolumn{1}{c|}{$8$}      & \multicolumn{1}{c|}{$80$}      &   $2240$    \\ \hline
\multicolumn{1}{|l|}{Deflation} & \multicolumn{1}{c|}{$15$}      & \multicolumn{1}{c|}{$5355$}      &  $50868675$     & \multicolumn{1}{c|}{$40$}      & \multicolumn{1}{c|}{$298480$}      & $494845859200$      \\ \hline 
\end{tabular}
\vspace{5pt}
\caption{Number of possible prefix codes when $t=2$ and $k'=1$.}
\label{tab:Poss_t2k1}
\end{table}

\begin{table}[]
\centering
\begin{tabular}{l|ccc|ccc|}
\cline{2-7}
                                       &                            & $p=2$                      &       &                            & $p=3$                      &       \\ \cline{2-7} 
                                       & \multicolumn{1}{c|}{$r=1$} & \multicolumn{1}{c|}{$r=2$} & $r=3$ & \multicolumn{1}{c|}{$r=1$} & \multicolumn{1}{c|}{$r=2$} & $r=3$ \\ \hline
\multicolumn{1}{|l|}{Punc. and short.}        & \multicolumn{1}{c|}{$45$}      & \multicolumn{1}{c|}{$6885$}      & $14768325$        & \multicolumn{1}{c|}{$120$}      & \multicolumn{1}{c|}{$275520$}      &  $49075622400$     \\ \hline
\multicolumn{1}{|l|}{Deflation} & \multicolumn{1}{c|}{$315$}      & \multicolumn{1}{c|}{$213648435$}      &   $4.89 \cdot 10^{17}$   & \multicolumn{1}{c|}{$3640$}      & \multicolumn{1}{c|}{$4503097318720$}      &  $2.80\cdot 10^{27}$     \\ \hline
\end{tabular}
\vspace{5pt}
\caption{Number of possible prefix codes when $t=3$ and $k'=1$.}
\label{tab:Poss_t3k1}
\end{table}

\section{How to construct codes with improved parameters}
So far we have relied only on a lower bound on the minimum distance of the deflated code. But, as Example \ref{ex:ImprovedDistance} shows,
a clever choice of $S_q'$ might lead to a minimum distance strictly larger than this lower bound.
Inspired by the classical idea of hitting sets \cite{GrasslWhite2004}, the work \cite{Our_puncture} describes conditions under which the bound on the minimum distance from puncturing can be improved. We want to generalize these ideas to deflation. We start by defining the set
\begin{align*}
    \mathcal{M}_{t,A}=\{\symp{\mathbf{a},\mathbf{u}}{\mathbf{b},\mathbf{v}}\in A\mid w_s\symp{\mathbf{u}}{\mathbf{v}}\leq d-t\}.
\end{align*}
First, we remark that $\mathcal{M}_{t,S_q^{\perp_s}\setminus S_q}$ will consist only of elements with symplectic weight equal to $d$, but it does not necessarily contain all of these. Namely, it contains only those that also satisfy $w_s\symp{\mathbf{a}}{\mathbf{b}}=t$.
Our motivation for looking at this set is that it contains some of the vectors that are `problematic' if we want a minimum distance higher than $d-t$ after deflating.
Note that we need to ensure that these vectors do not have prefix in $S_q'^{\perp_s}$ because they would otherwise be part of $(S_q^{S_q'})^{\perp}\setminus S_q^{S_q'}$ due to Proposition \ref{prop:dual}. 

The other `problematic' vectors are low-weight vectors of $S_q$ that are moved to the dual during deflation. Potential such vectors are exactly $\mathcal{M}_{t,S_q}$, and this set can contain vectors of symplectic weight less than $d$ if the code is impure. In order to deal with these vectors, we need to ensure that the prefix of such a vector either lies in $S_q$, meaning that it is kept in the stabilizer of the deflated code, or is outside $S_q'^{\perp_s}$, meaning that it is removed. Thus, we must require that $\pi_{\bar{I}}(\mathcal{M}_{t,S_q})\cap S_q'^{\perp_s}\setminus S_q'=\emptyset$ since we would otherwise have an element of weight less than or equal to $d-t$ in the deflated code. 

To summarize, if we can choose $S_q'$ such that 
\begin{align*}
  \big(\pi_{\bar{I}}(\mathcal{M}_{t,S_q^{\perp_s}\setminus S_q})\cap S_q'^{\perp_s}\big)
  \cup
  \big(\pi_{\bar{I}}(\mathcal{M}_{t,S_q})\cap S_q'^{\perp_s}\setminus S_q'\big)=\emptyset,
\end{align*}
we will have that the minimum distance of the deflated code is at least $d-t+1$. 

\section{Relation to classical codes}\label{sec:deflationClassical}
Since both puncturing and shortening are well-known techniques from classical error-correction, a natural question would be if a similar generalization of these techniques exist in the classical setting as well. In this section, we show that this is indeed possible, but explain why a classical deflation is unlikely to be as useful as its quantum counterpart.
We assume familiarity with standard concepts from classical coding theory, which can be found in standard textbooks such as~\cite{huffman2003fundamentals,MacWilliamsSloane}.

Let $C$ be a linear $[n,k,d]_q$ code, and let $C'$ be a linear $[t,k']_q$ code.
Following the same idea as in Section~\ref{sec:deflation}, the deflated code of $C$ is given by
\begin{align*}
    C^{C',I}=\tilde{\pi}_I(\mathrm{ker}(\tilde{\sigma}_{B_{C'},I})),
\end{align*}
where $\tilde{\pi}$ and $\tilde{\sigma}_{B_{C'},I}$ are defined similarly to~\eqref{eq:pi_I} and \eqref{eq:sigma}.
More precisely, $\tilde{\pi}$ removes all the entries with indices in $I$, and $\tilde{\sigma}_{B_{C'},I}$ maps a vector to its inner products with all elements of a basis for $C'^\perp$.
In this way, $C^{C',I}$ is obtained by picking out codewords whose prefix is in $C'$, before removing the entries that make up these prefixes. This corresponds exactly to the procedure performed on the quantum stabilizer.

In general, however, it is not easy to determine the dimension of this deflated code.
Namely, the dimension is given by
\begin{equation*}
  \dim C^{C',I} = k - \big(\dim\tilde{\pi}_{\bar{I}}(C) - \dim(\tilde{\pi}_{\bar{I}}(C)\cap C')\big).
\end{equation*}
If $I$ is contained in an information set (i.e. the columns of a generator matrix of $C$ corresponding to $I$ is a linearly independent set), then $\tilde{\pi}_{\bar{I}}(C)$ has dimension $\lvert I\rvert=t$, and the dimension therefore reduces to $\dim C^{C',I} = k+k'-t$.
Thus, the deflated code has parameters $[n-t,k-(t-k')]_q$.
One simple way to ensure that the above always happens, is to assume that $C^\perp$ has minimum distance larger than $t$.
Since the minimum distance of $C^{C',I}$ is bounded below by $d-t$ as in the quantum setting, the codes where we can directly control both the dimension and the minimum distance after deflation are restricted to those $C$ and $C'$ that satisfy $d(C^\perp)< t < d(C)$. 

Furthermore, we remark that there is a fundamental difference between the classical and quantum constructions.
In the quantum case, the dimension is increased, typically at the cost of reducing the minimum distance.
In the classical case, the dimension cannot increase -- and will in fact decrease in most cases -- but we will still tend to see a decrease in minimum distance.
In conclusion, it seems difficult to control the parameters of deflated codes in the classical setting unless we have particular assumptions on the minimum distance of the code and its dual. Even in this case, both the dimension and minimum distance seem to suffer, suggesting that the resulting deflated codes are unlikely to be better than the original classical codes. We leave it as an open problem to consider if any further assumptions can improve the bounds on the parameters for classical codes and in that case make deflation a useful tool here as well.

\section{Conclusion}
In this work, we have generalized and unified the puncturing and shortening techniques for quantum stabilizer codes.
We have introduced the notion of deflating stabilizer codes, which has both shortening and puncturing as special cases. In Theorems~\ref{thm:parametersPure} and \ref{thm:parameters_impure}, we determine the parameters for the deflated codes. Especially, in Theorem~\ref{thm:parameters_impure} we also relax the condition that the code needs to be pure, allowing, e.g., shortening of impure codes with certain properties.
Furthermore, we investigated the freedom gained in the choice of prefixes when generalizing to deflated codes, and we have illustrated by an example that this may lead to better parameters than sequential application of just shortening and puncturing.
As we show in Theorems~\ref{thm:parametersPure} and \ref{thm:parameters_impure}, the minimum distance obeys the same bound as seen in puncturing and shortening, namely $\tilde{d}\geq d-t$. But a procedure similar to the one used in~\cite{Our_puncture} allows setting up criteria that ensure that the deflated code satisfies $\tilde{d}\geq d-t+1$.
In particular, $S_q'$ is required to be chosen in a specific manner, and we leave it as an open question, whether codes satisfying this criterion exist, and whether they can be used to construct quantum codes with previously unrealizable parameters.

\section*{Acknowledgments}
This work was supported, in part, by the Danish National Research Foundation (DNRF), through the Center of Excellence CLASSIQUE, grant nr. DNRF187, and, in part, by the Poul Due Jensen's foundation via the grant SWIFT.

\bibliographystyle{IEEEtran}
\bibliography{references}

\end{document}